\documentclass{article}
\usepackage{ijcai22}

\usepackage[colorlinks]{hyperref}
\usepackage{times}
\usepackage{soul}
\usepackage{url}
\usepackage{mathrsfs} 
\usepackage{graphicx}
\usepackage{multirow}
\usepackage{amsmath}
\usepackage{amsthm}
\usepackage{booktabs}
\usepackage{algorithm}
\usepackage{algorithmic}
\usepackage{xcolor}
\usepackage{subfigure}
\usepackage{amsfonts}
\def\ie{\textit{i.e.}}
 
\def\eg{\textit{e.g.}}

\usepackage{afterpage}

\usepackage{caption}
\DeclareMathOperator*{\argmin}{arg\,min}

\title{Shielding Federated Learning: Robust Aggregation with Adaptive Client Selection}

\author{
Wei Wan$^{1,3,4,5}$\and
Shengshan Hu$^{1,3,4,5}$\and
Jianrong Lu$^{1,3,4,5}$\and
Leo Yu Zhang$^7$\and\\
Hai Jin$^{2,3,4,6}$\And
Yuanyuan He$^{1}$\footnote{Corresponding Author}\
\affiliations
$^1$School of Cyber Science and Engineering, Huazhong University of Science and Technology\\
$^2$School of Computer Science and Technology, Huazhong University of Science and Technology\\
$^3$National Engineering Research Center for Big Data Technology and System\\
$^4$Services Computing Technology and System Lab\\
$^5$Hubei Engineering Research Center on Big Data Security\\
$^6$Cluster and Grid Computing Lab \\
$^7$School of Information Technology, Deakin University
\emails
\{wanwei\_0303, hushengshan, lujianrong, hjin, yuanyuan\_cse\}@hust.edu.cn,\\
leo.zhang@deakin.edu.au
}

\begin{document}
\maketitle
\renewcommand{\thefootnote}{}
\footnotetext{Published on \href{https://www.ijcai.org/proceedings/2022/0106.pdf}{IJCAI'22}
}
\begin{abstract}
\textit{Federated learning} (FL) enables multiple clients to collaboratively train an accurate global model while protecting clients' data privacy. However, FL is susceptible to Byzantine attacks from malicious participants. Although the problem has gained significant attention, existing defenses have several flaws:  the server irrationally chooses malicious clients for aggregation even after they have been detected in previous rounds; the defenses perform ineffectively against sybil attacks or in the heterogeneous data setting.

To overcome these issues, we propose \href{https://github.com/hustweiwan/MAB-RFL}{MAB-RFL}, a new method for robust aggregation in FL. By modelling the client selection as an extended \textit{multi-armed bandit} (MAB) problem, we propose an adaptive client selection strategy to choose honest clients that are more likely to contribute high-quality updates. We then propose two approaches to identify malicious updates from sybil and non-sybil attacks, based on which rewards for each client selection decision can be accurately evaluated to discourage malicious behaviors. MAB-RFL achieves a satisfying balance between exploration and exploitation on the potential benign clients. Extensive experimental results show that MAB-RFL outperforms existing defenses in three attack scenarios under different percentages of attackers.

\end{abstract}

\section{Introduction}
\textit{Federated learning} (FL)~\cite{FedAvg} is a newly emerging machine learning paradigm that aims to train a shared global model on decentralized data with user's privacy guarantee. A FL system consists of a central server and multiple clients. The clients do not share their raw data with the central server, instead, each client trains a local model via fine-tuning the global model over its local data set and then sends the local model update (\ie, the difference between local model and the global model) to the server. The server aggregates all the local updates to obtain a new global model and shares it with the clients for next iteration. FL could not only construct a model with high accuracy but also address the privacy concerns, thus drawing extensive research attention recently.

However, FL is susceptible to Byzantine attacks from malicious clients due to the distributed nature. For example, attackers can upload poisoned local updates via contaminating the training data (\ie, data poisoning attacks~\cite{howtobackdoor,PCA}) or tampering with the local model updates directly (\ie, model poisoning attacks~\cite{ALittleIsEnough,AGRTailored,modelpoioningattack}), in order to damage the global model or even leave a backdoor.

To mitigate Byzantine attacks, plenty of defensive schemes have been proposed. Their key ideas include eliminating updates that are far from the overall distribution through similarity detection~\cite{Krum,FABA},  circumventing the malicious parameters through statistical characteristics~\cite{TrimmedMean,GeoMed,SLSGD}, detecting outliers based on the extracted features of updates~\cite{PCA,AGRTailored}, relying a clean dataset to detect abnormality~\cite{Zeno,FLTrust}, and directly clipping the local updates with large magnitudes~\cite{CC}.

Albeit tremendous in types, all the existing defenses suffer from one or multiple  limitations as follows. Firstly, they paid little attention to the client selection step. In each iteration, the server is set to randomly select a subset of clients~\cite{FLTrust} or simply choose all of them to participate in the aggregation~\cite{Krum,FABA,CC}. Second, they cannot effectively defend against sybil attacks, in which all the malicious updates are similar or even identical~\cite{ALittleIsEnough,AGRTailored,modelpoioningattack}. Third, they perform badly in non-IID (\textit{independently identically distribution}) scenarios, where all the updates are dramatically different from each other~\cite{FLTrust,DiverseFL}. Lastly, they assume that the server holds a validation dataset that has the same distribution with the  training set of clients~\cite{FLTrust,Zeno,DiverseFL}, which obviously violates the privacy requirements of FL.

To address these issues, we propose a new multi-armed bandit based robust FL scheme, named MAB-RFL. Our scheme models the client selection process in federated learning as an extended MAB problem enabling the server to adaptively select updates that are more likely to be benign. Before applying existing MAB solutions to FL, there are two key challenges needed to be addressed. First, in standard MAB problem, only one arm is chosen at each time step, however, in FL scenarios, a mass of clients are selected to guarantee the high accuracy of the global model. Inspired by the Thompson sampling~\cite{thompsonsampling}, we propose an adaptive client selection strategy to decide the current round participants. Second, in the standard MAB scenario, the reward for each decision comes from the slot machine directly once an arm is played. Nevertheless, in FL the server has no idea how to evaluate each decision of client selection. Therefore, we propose two approaches to identify malicious updates from different attacks (\ie, sybil and non-sybil), based on which rewards can be easily computed. 
Note that MAB methods have been applied in FL in other works, but they focus on reducing communication overheads~\cite{mab_efficient2} or constructing high-quality models~\cite{mab_hightqualitymodel}, it is still unclear how to apply MAB algorithms to detect Byzantine attacks.


In summary, we make the following contributions:
\begin{itemize}
    \item  We model the client selection in Byzantine settings as a MAB problem and propose a novel adaptive client selection strategy, enabling the server to rationally select participants that are more likely to be honest and useful.
   
    \item We propose two new solutions to identify malicious clients from sybil and non-sybil attacks, such that the rewards for client selection decisions can be accurately evaluated to discourage misbehaviour. 

    \item We conduct extensive experiments to evaluate MAB-RFL. The results show that the proposed scheme is effective for three attack scenarios under different percentages of attackers.
    
\end{itemize}
\section{Background}
\subsection{Federated Learning}
We consider a general FL system, consisting of a central server and $K$ clients. Each client $k$ ($k=1,2,...,K$) has a dataset $D_{k}$, the size of which is denoted as  $|D_{k}|=n_{k}$. It is worth noting that each local dataset may be subject to a different distribution, that is, the clients' data may be distributed in a non-IID  way. The clients aim to collaboratively train a shared global model $w$. Apparently, the problem can be solved via minimizing the empirical loss, \ie, $\argmin_{w} f(D,w)$, where $D=\bigcup_{k=1}^{K} D_{k}$ and $f(D,w)$ is a loss function (\eg, mean absolute error, cross‐entropy). However, the optimization requires all the clients to share their raw data to a central server, which would result in a serious threat to client's privacy. Instead, FL obtains $w$ by optimizing $\argmin_{w} \sum_{k=1}^k f(D_{k},w)$. Specifically, the FL system iteratively performs the following three steps until the global model converges:
\begin{itemize}
	\item \textbf{Step I:} In the $t$-th iteration, the central server broadcasts a global model $w_{t}$ to the clients;
	\item \textbf{Step II:} After receiving $w_{t}$, each client $k$ trains a new local model $w_{t}^{k}$ over $D_{k}$ by solving the optimization problem $\argmin_{w_{t}^k} f(D_{k},w_{t}^k)$ and then uploads the local model update $g_{t}^k := w_{t}^k - w_{t}$ to the server;
	\item \textbf{Step III:} The server aggregates all the local updates according to client's proportional dataset size as follow:
	\begin{equation}
	w_{t+1}\gets w_{t}+\sum_{k=1}^{K} \frac{n_{k}}{n} g_{t}^{k}, \text{where} \ n=\sum_{k=1}^{K}n_{k}.
	\end{equation}
\end{itemize}

\subsection{Multi-Armed Bandit Problem and Thompson Sampling}
The \textit{multi-armed bandit} (MAB) problem is a classical framework which studies the exploration/exploitation trade-off in sequential decision problems. Specifically, assuming that we are given a slot machine with $K$ arms, at each time step $t = 1, 2, 3, ..., T$, we play an arm $k(t)\in\{1,2,...,K\}$ and observe the corresponding reward $R_{k(t)}$. In this work we consider the Bernoulli bandit, \ie, $R_{k(t)}$ is either 0 or 1. The MAB aims to minimize the expected total regret over $T$ rounds, where the regret is the amount we lose because of not playing optimal arm in each round. Formally, the $T$
period expected cumulative regret is given by:
\begin{equation}
\begin{aligned}
	\mathbb{E}[Regret(T)] &= TR_{k^{*}}-\mathbb{E}[\sum_{t=1}^{T}R_{k(t)}]=\sum_{t=1}^{T} \textbf{1}\{k_{t}\neq k^{*}\},
	\end{aligned}
\end{equation}
where $k^{*}$ denotes the best arm (\ie, $R_{k^{*}} = \max_{k}R_{k}$).
The Thompson sampling algorithm~\cite{thompsonsampling} is a popular solution to the MAB problem which models the mean reward of each arm using a Beta distribution and then chooses the arm with the largest mean reward.

{\color{black}Alg.~\ref{alg:TS} provides the details of Thompson sampling for the Bernoulli bandit. Note that $S_{k}$ and $F_{k}$ denote the number of successes and failures outcome of the arm $k$ respectively, and both of them are initialized to 1, because Beta$(1,1)$ is the uniform distribution on $(0,1)$, which means that each arm has the same probability of being chosen at the beginning.}

\begin{algorithm}[tb]
\caption{Thompson sampling for the Bernoulli bandit}
\label{alg:TS}
\begin{algorithmic}[1] 
\FOR{$k=1,2,...,K$}
    \STATE $S_{k}=\gets 1, F_{k}\gets 1$.
\ENDFOR
\FOR{$t=1,2,...,T$}
    \FOR{$k=1,2,...,K$}
      \STATE Draw $\hat{R_{k}}$ according to Beta($S_{k},F_{k}$).
    \ENDFOR
    \STATE Draw arm $k(t) \gets \arg\max_{k}\hat{R_{k}}$ and observe reward $R_{k(t)}$.
    \IF{$R_{k(t)}=1$}
        \STATE $S_{k(t)} \gets S_{k(t)}+1$.
    \ELSE
        \STATE $F_{k(t)} \gets F_{k(t)}+1$.
    \ENDIF
\ENDFOR
\end{algorithmic}
\end{algorithm}

\section{Problem Setup}\label{Problem Setup}
\subsection{Attack Model}
We employ a similar attack model in previous works~\cite{FLTrust,DiverseFL}. To be specific, an adversary controls several clients (less than half of all participants), and it can poison the local data and updates of the controlled clients arbitrarily. However, the central server and benign clients are not under the control of the adversary. The only way the adversary can degrade the performance of the global model is by uploading poisoned local updates through compromised clients. We consider three kinds of attacks according to the adversary's knowledge about the FL system:

\textit{Agnostic attack:} the adversary knows neither benign updates nor the aggregation algorithm.

\textit{Partial knowledge attack:} the adversary only has the knowledge of benign updates.

\textit{Full knowledge attack:} the adversary knows both benign updates and aggregation algorithm.

Although the agnostic attack is the weakest type, it is the most realistic attack.
Partial knowledge attack and full knowledge attack are too powerful to be applied in practice, but they have been widely investigated in previous studies~\cite{FLTrust,AGRTailored,CC,momentum} to fully understand the threats of Byzantine attacks.

\subsection{Defense Model}
\label{Sec3.2: Defenes Model}
Our defense is deployed on the central server, and does not require any privacy information about participants. Specifically, the central server does not access the raw local training data, and is unknown about the number of compromised clients and even the training dataset size of each client.
Our defense is expected to achieve the following three goals simultaneously:
\begin{itemize}
    \item \textbf{Rationality.}
    The method should avoid choosing clients who continually uploaded low-quality updates in previous iterations. Note that the low-quality updates may be caused by adversary's meticulous design~\cite{AGRTailored,modelpoioningattack} or benign participants' poor training data ~\cite{weightattack}. 
   
    \item \textbf{Robustness.}
    The method should prevent malicious updates from degrading the accuracy or the convergence rate of the global model in all kinds of attack scenarios, \ie, the agnostic attack, the partial knowledge attack, and the full knowledge attack.
   
    \item \textbf{Fidelity.} 
    The method should not incur a drop in non-adversarial scenarios. In other words, the defense should be as accurate as FedAvg when there is no attack.
    
\end{itemize}

{\color{black}To the best of our knowledge, none of existing Byzantine-robust schemes satisfy the rationality goal. They sample either a random subset of clients or all participants in each iteration. Besides, some schemes such as Krum, Median, FABA, and \textit{Centered Clip} (CC) violate the robustness goal, because they perform poorly in full-knowledge settings. Moreover, Krum does not satisfy the fidelity goal.}

\begin{figure}[t]
	\centering
	\subfigure[Evaluation of sampling]{\label{fig:sampling}
		\includegraphics[width=0.48\columnwidth]{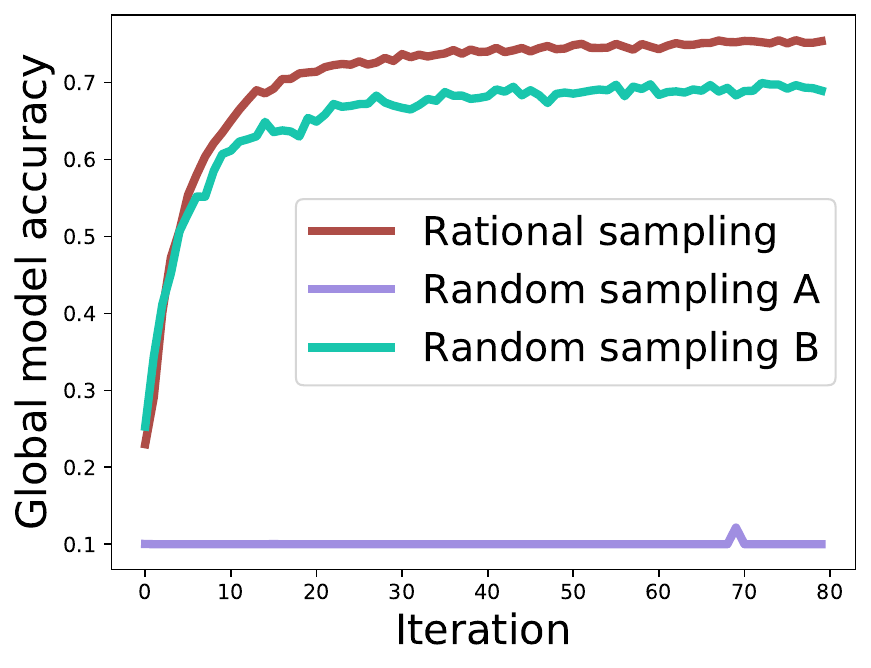}}
	\subfigure[Evaluation of similarity]{\label{fig:max_cos_similarity}
		\includegraphics[width=0.48\columnwidth]{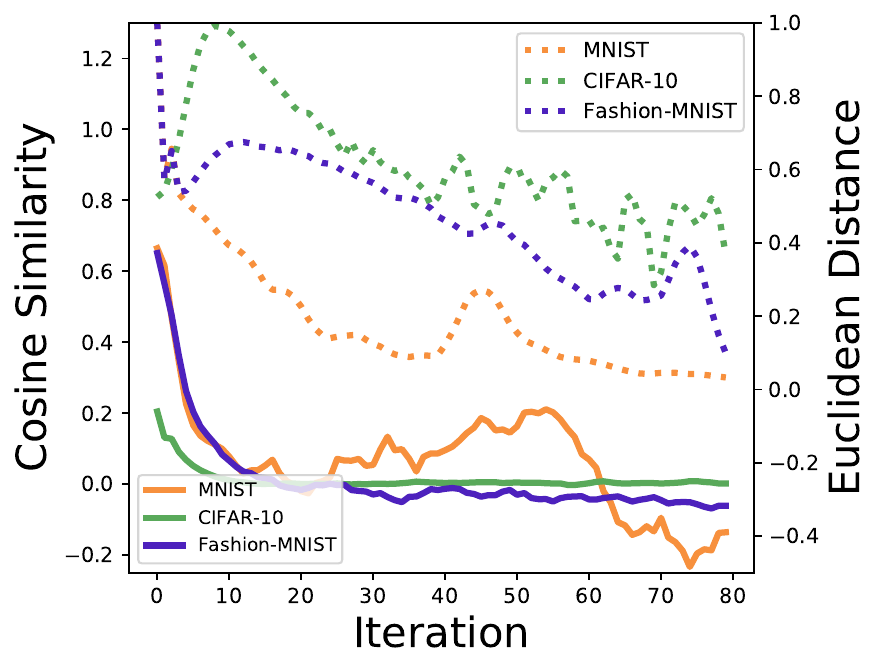}}
	\caption{(a) The comparison between random sampling and rational sampling; (b) Maximum cosine similarity between benign updates.}
	\label{fig:lf}
\end{figure}

\begin{algorithm}[t]
\caption{A Complete Description of MAB-RFL}
\label{alg:MAB-RFL}
\textbf{Input}: $K$ clients; estimated maximum cosine similarity $c_{max}$ and minimum cosine similarity $c_{min}$; importance of historical information $\lambda$; acceptable difference between clusters $\alpha$. \\
\textbf{Output}: Global model $w_{t+1}.$
\begin{algorithmic}[1] 
\STATE \textbf{Initialization:} $B_{k}\gets1, M_{k}\gets1, \forall k \in [K]$.
\STATE //Step I: Select reliable clients.
\STATE $L_{selected} \gets ACS(\{B_{k},k\in[K]\}, \{M_{k},k\in[K]\})$
\STATE Broadcast the current global model $w_{t}$ to the clients in $L_{selected}$ and gather the corresponding local updates $g = \{g_{k}, k \in L_{selected}\}$.
\STATE //Step II: Identify sybil attacks.
\STATE $L_{sybil} \gets ISA(L_{selected},g)$.
\STATE $L_{selected} \gets L_{selected}-L_{sybil}$.
\STATE $g \gets \{g_{k}, k \in L_{selected}\}$.
\STATE //Step II: Identify non-sybil attacks.
\STATE  $L_{nsybil}\gets INSA(L_{selected},g)$.
\STATE $L_{selected} \gets L_{selected}-L_{nsybil}$.
\STATE //Step III: Distribute rewards.
\STATE $M_{k} \gets M_{k}+1, \forall k \in L_{sybil}\cup L_{nsybil}$.
\STATE $B_{k} \gets B_{k}+1, \forall k \in L_{selected}$.
\STATE //Step IV: Aggregate updates.
\STATE Calculate the learning rate: \\
$\eta_{t} \gets Mean(\{||g_{k}||, k \in L_{selected}\})$.
\STATE Calculate the normalized momentum vectors:\\
$mm \gets \{\overline{m_{t}^{k}}, k \in L_{selected}\}$ through Eq.~(\ref{new_momentum}) and Eq.~(\ref{normalize}).
\STATE $w_{t+1} \gets w_{t}+ \eta_{t} \cdot Mean(mm)$.
\RETURN $w_{t+1}$
\end{algorithmic}
\end{algorithm}

\section{MAB-RFL: Multi-Armed Bandit Based Robust Federated Learning}
\subsection{Motivation Behind Rational Client Selection}
\label{Sec4.1: Motivation}

After reviewing the existing defenses against Byzantine attacks in FL, we find that they choose either all clients~\cite{FABA,Krum,CC} or a random subset of them~\cite{FLTrust}. 
This irrational client selection strategy may continually select malicious clients to participate in the aggregation, even after they have been identified in previous iterations, leading to a slow convergence and a waste of communication and computation resources. Fig.~\ref{fig:sampling} gives a toy example. We consider a FL system containing $50$ clients, among which $15$ are malicious, and $10$ clients will be selected at each iteration. Random sampling $A$ randomly selects $10$ clients and aggregates all the updates directly. Random sampling $B$ also randomly selects $10$ clients but discards all the malicious updates before aggregation. Rational sampling samples 10 benign clients and aggregates all the updates directly.
The results show that random sampling $B$ converges slower and the accuracy of the final global model is about $5\%$ lower compared with the rational sampling.
Therefore, a rational server should always choose honest clients who are more likely to submit  high-quality updates.


\subsection{A Complete Description of MAB-RFL}
Alg.~\ref{alg:MAB-RFL} shows our complete MAB-RFL scheme.
In Step I (Lines 2-4), the server adaptively selects clients and broadcasts the current global model to the clients. In Step II (Lines 5-11), the server identifies sybil updates (Lines 5-8) as well as non-sybil but malicious updates (Lines 9-11). In Step III (Lines 12-14), the server assigns reward to each selected client. In Step IV (Lines 15-19), the server updates the global model based on the normalized momentum vectors of the remaining clients. Note that we set the learning rate as the average of the magnitudes of all the left updates (Line 16) to make up for the slow or fast learning caused by the normalization.
Next, we present details for the first three steps.



\begin{figure*}[t]
\centerline{\includegraphics[width=1.8\columnwidth]{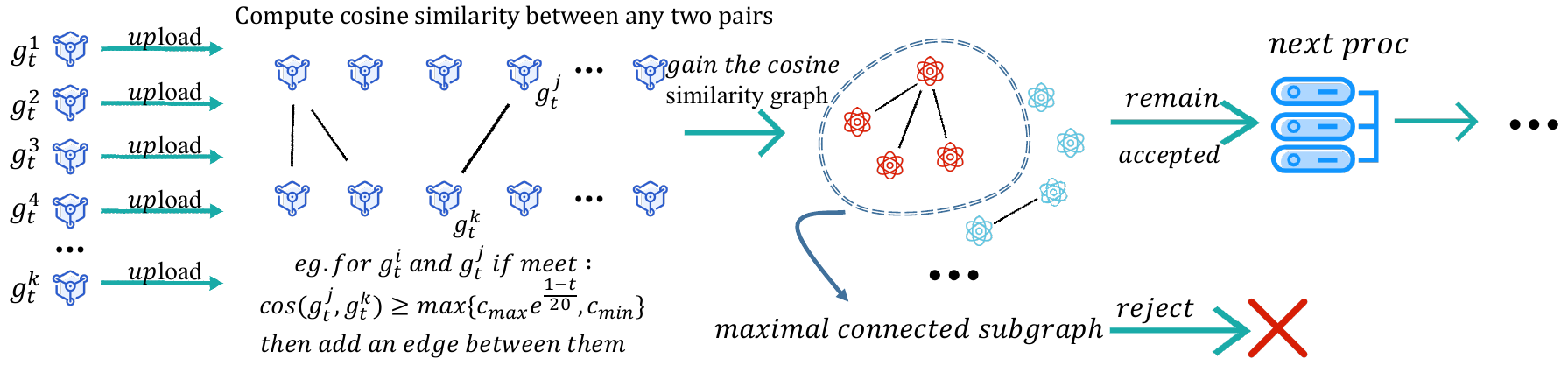}}
\caption{Illustration of sybil attack detection}
\label{fig:sybil_detection}
\end{figure*}

\begin{algorithm}[t]
\caption{ACS (Adaptive Client Selection)}
\label{alg:client selection}
\textbf{Input}: The frequencies that the local updates from each client $k \in [K]$ are regarded as benign $B_{k}$ and malicious $M_{k}$.\\
\textbf{Output}: The selected clients list $L_{selected}$.

\begin{algorithmic}[1] 
\FOR{$k=1,2,...,K$}
    \STATE $p_{k}$ $\gets$ Beta($B_{k},M_{k}$).
    \STATE Add $k$ to the selected clients list $L_{selected}$ at the probability $p_{k}$.
\ENDFOR
\IF {$L_{selected}$ = $\emptyset$}
\STATE Assign a random nonempty subset of $[K]$ to $L_{selected}$.
\ENDIF
\RETURN $L_{selected}$
\end{algorithmic}
\end{algorithm}

\subsection{Adaptive Client Selection}
In the standard Thompson sampling algorithm, only the arm with the maximal mean reward (estimated by a Beta distribution) will be played at each time step. An  intuitive idea for extending it to FL is to select clients with top-$m$ ($m$ is the number of benign clients) mean rewards. However, it relies on the assumption that the number of benign clients is fixed and known to the server.  This assumption is unrealistic since FL is a dynamic distributed network where honest and compromised clients can drop out arbitrarily. 

To get rid of the dependence on $m$, we propose selecting each client with a changing probability {\color{black}(\ie, the probability varies with each client’s past performance, rather than using a fixed probability)} as shown in Alg.~\ref{alg:client selection}. Specifically, we set the  probability of each client $k$ being selected to be $p_{k}$, which is drawn from Beta($B_{k}$, $M_{k}$) distribution (Lines 1-4). Here $B_{k}$ and $M_{k}$ denote the frequency that the local updates from client $k$ are identified as benign and malicious (through our methods described in Section~\ref{sec:clientIden}), respectively. Initially, we set $B_{k} = M_{k} = 1$. If none of the clients are selected in a round, we will randomly choose a subset of clients (Lines 5-8).

This method can achieve a well balance between exploration and exploitation. A client who consistently uploads benign updates (\ie, a large $B_{k}$ and a small $M_{k}$) will obtain a large $p_{k}$ from Beta($B_{k}$, $M_{k}$), and it is easier to be selected (exploitation), while a rarely selected client (\ie, both $B_{k}$ and $M_{k}$ are small) may also get a large $p_{k}$ (exploration), because the Beta distribution has a large variance when the parameters are small.

\begin{algorithm}[t]
\caption{ISA (Identifying Sybil Attacks)}
\label{alg:sybil detection}
\textbf{Input}: The selected clients list $L_{selected}$ and the corresponding local updates $\{g_{k}, k\in L_{selected}\}$. \\
\textbf{Output}: Rejected clients list $L_{sybil}$.

\begin{algorithmic}[1] 
\STATE Construct a graph $G=(V,E)$, where the vertex set $V=L_{selected}$, and the edge set $E$ is obtained through Eq.~(\ref{edge construct}).
\STATE Assign the maximal connected subgraph of G to $G^{'}=(V^{'},E^{'})$.
\STATE $L_{sybil}\gets V^{'}$.
\RETURN $L_{sybil}$
\end{algorithmic}
\end{algorithm}

\subsection{Abnormal Clients Identification}
\label{sec:clientIden}

\paragraph{Identifying Sybil Attacks.}

Sybil attack is a recently proposed special Byzantine attack method~\cite{Sybils}, where all the compromised clients upload totally identical local updates to launch covert attacks.  This kind of attack poses a significant threat to existing defenses because existing defenses assume that benign updates are much similar and discard updates that are far from the overall distribution.

Before presenting our defensive method against sybil attacks, we introduce a basic fact that has been long misunderstood in the literature. In FL,  the similarity between benign updates is as high as we expected when we use Euclidean distance as the evaluation metric. If we use cosine similarity as the evaluation metric, however, the benign updates have low similarity even in IID setting. This phenomenon is also observed by our experiments shown in Fig.~\ref{fig:max_cos_similarity}, where the training data among clients are IID  and each client has the same dataset size. We can see that after several iterations, the maximum cosine similarity  reduces sharply to 0 (indicating a larger dissimilarity), while the Euclidean distances (computed after normalization) get smaller (indicating a larger similarity).
We owe this phenomenon to the fact that when the model converges, benign gradients will gradually become zero, thus making their cosine similarities close to zero.



Based on the findings above, we propose mitigating sybil attacks by using undirected graph. Specifically, we regard each uploaded local update as a vertex and construct a graph $G=(V,E)$ through the following rule: if two updates are too similar in direction (in terms of cosine similarity), an undirected edge will be added. Formally, an undirected edge is added between vertex $j$ and $k$ if and only if the updates $g_{t}^{j}$ and $g_{t}^{k}$ satisfy the following property:
\begin{equation}
    cos(g_{t}^{j},g_{t}^{k}) \geq \max\{c_{max}e^\frac{{1-t}}{20}, c_{min}\},
    \label{edge construct}
\end{equation}
where $cos$ denotes the cosine similarity, $c_{max}$ and $c_{min}$ $(-1 < c_{min} < c_{max} <1)$ represent the estimated maximum and minimum cosine similarities,  and  $t$ is the current iteration. {\color{black}Note that the tolerable cosine similarity (right of the above inequality) decreases as iterations go on, which indicates that it is harder for the adversary to launch covert sybil attack.} 

After constructing the graph, the updates (vertexes) in the maximal connected subgraph $G^{'}=(V^{'},E^{'})$ of G will be considered as malicious ones. 
Fig.~\ref{fig:sybil_detection} gives a simple illustration for this process and Alg.~\ref{alg:sybil detection} shows the details.

\begin{algorithm}[t]
\caption{INSA (Indentifying Non-Sybil Attacks)}
\label{alg:abnormal behavior detection}
\textbf{Input}: The selected clients list $L_{selected}$ and the corresponding local updates $\{g_{k}, k\in L_{selected}\}$. \\
\textbf{Output}: Rejected clients list $L_{nsybil}$.

\begin{algorithmic}[1] 
\FORALL{$k \in L_{selected}$}
    \STATE Calculate the momentum vector $m_{t}^{k}$ according to Eq.~(\ref{new_momentum}) and normalize $m_{t}^{k}$ to be $\overline{m_{t}^{k}}$ through Eq.~(\ref{normalize}).
\ENDFOR
\STATE Extract the key features of each $\overline{m_{t}^{k}}$ by PCA.
\STATE Divide $L_{selected}$ into a larger cluster $C_{l}$ and a smaller cluster $C_{s}$ by applying agglomerative clustering algorithm to the key features.
\STATE Calculate the mean momentum vector of each cluster:\\
$m_{l} \gets Mean(\{m_{t}^{k}, k \in C_{l}\})$;\\
$m_{s} \gets Mean(\{m_{t}^{k}, k \in C_{s}\})$.
\IF{$cos(m_{l},m_{s})\textgreater \alpha$}
        \STATE $L_{nsybil} \gets \emptyset$.
\ELSE
        \STATE $L_{nsybil} \gets C_{s}$.
\ENDIF
\RETURN $L_{nsybil}$
\end{algorithmic}
\end{algorithm}

\paragraph{Identifying Non-sybil Attacks.}

In non-sybil settings where malicious updates obviously differ from each other, it becomes difficult to identify them. A pivotal reason for the difficulty  lies in the extremely high dimension of local models, which contains deciding features as well as redundant features that will mask the differences between benign and malicious updates. Moreover, the non-IID property of FL naturally incurs disparity among benign updates, making it more difficult to find out abnormal ones. 
In light of this, we propose squeezing the feature space before filtering out updates. In order to counter the non-IID issue, we further make use of Momentum~\cite{momentum} to reduce the variance between updates such that an IID-like setting can be built.
Specifically, our approach contains three steps as shown in Alg.~\ref{alg:abnormal behavior detection}: 1)  calculating the momentum vector of each client (Lines 1-3); 2) using \textit{principal component analysis} (PCA) to extract the key features of each momentum vector (Line 4); 3) applying agglomerative clustering algorithm to divide the momentum vectors  and choosing the appropriate vectors for aggregation (Lines 5-12).

Traditionally, the momentum vector of client $k$ at iteration $t$ is given by:
\begin{equation}
     m_{t}^{k} = g_{t}^{k} + \lambda m_{t-1}^{k},
    \label{momentum}
\end{equation}
where $\lambda\in(0,1)$ indicates the importance of historical information, $g_{t}^{k}$ is the local update, and the initial momentum $m_{0}^{k}$ is generally set to be $0$. Nevertheless, it is not suitable for our case because clients are not selected continuously in FL.
We thus slightly revise the momentum vector calculation as:
\begin{equation}
    m_{t}^{k} = g_{t}^{k} + \lambda^{t-t_{k}} m_{t_{k}}^{k},
    \label{new_momentum}
\end{equation}
{\color{black}where $t_{k}$ is the latest iteration when client $k$ is selected. The intuition behind Eq.~(\ref{new_momentum}) is that the historical information $m_{t_{k}}^{k}$ of a chronically unselected client is obsolete, thus multiplied by a smaller discount factor. Note that Eq.~(\ref{new_momentum}) and Eq.~(\ref{momentum}) are equivalent without using client selection strategy (\ie, $t_{k} = t-1$).}

Besides, attackers may upload updates with large magnitudes to dominate the global update. We thus normalize the momentum vector as:
\begin{equation}
    \overline{m_{t}^{k}} =\frac{m_{t}^{k}}{||m_{t}^{k}||}.
    \label{normalize}
\end{equation}
This step ensures that a single malicious update has limited impact on the aggregated global update even though it could circumvent the detection.

In order to prevent MAB-RFL from discarding updates excessively, we  set a threshold $\alpha$ to determine if  all the momentum vectors will be used for aggregation (Lines 7-9).


\subsection{Reward Distribution Mechanism}
A correct reward distribution mechanism is important for MAB-RFL to reduce the possibility of selecting abnormal clients in subsequent iterations. As shown in Alg.~\ref{alg:MAB-RFL}, after identifying abnormal clients who conduct sybil and non-sybil attacks, they will get a reward 0 (\ie, $M_{k} \gets M_{k}+1$, Line 13) and be discarded in the current round (Line 7 and Line 11). The remaining selected clients get a reward 1 (\ie, $B_{k} \gets B_{k}+1$, Line 14) and their updates are used for final aggregation (Lines 15-19).

\begin{figure}[t]
	\centering
	\subfigure[MNIST]{\label{fig:lf_m}
		\includegraphics[width=0.48\columnwidth]{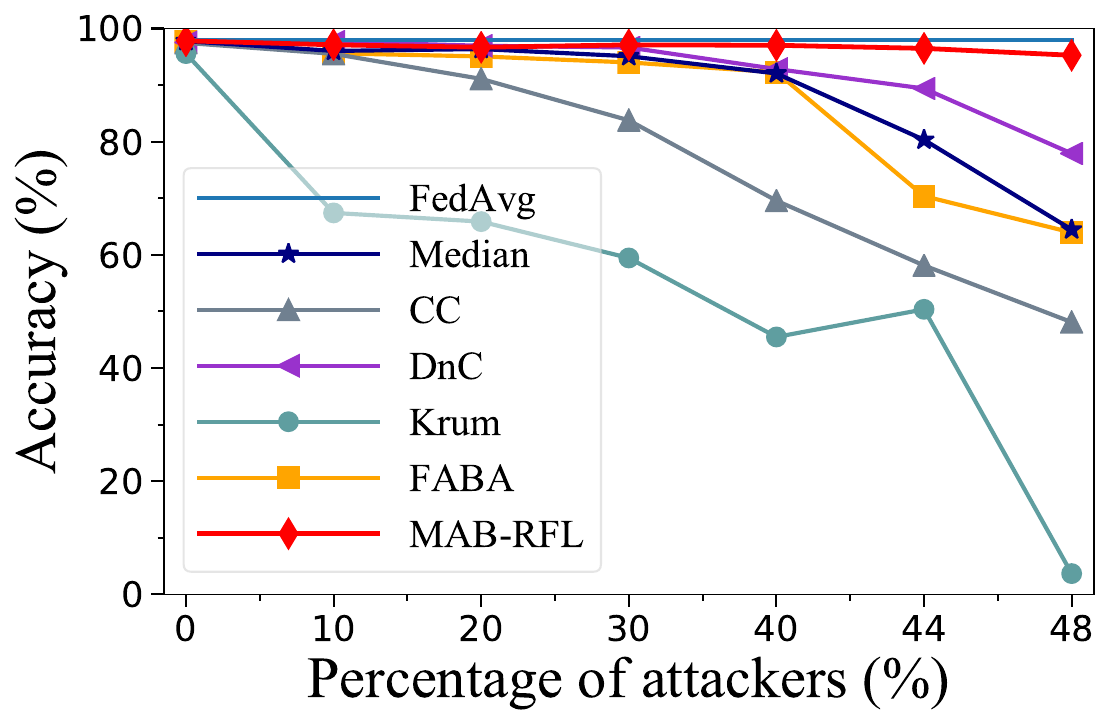}}
	\subfigure[CIFAR-10]{\label{fig:lf_c}
		\includegraphics[width=0.48\columnwidth]{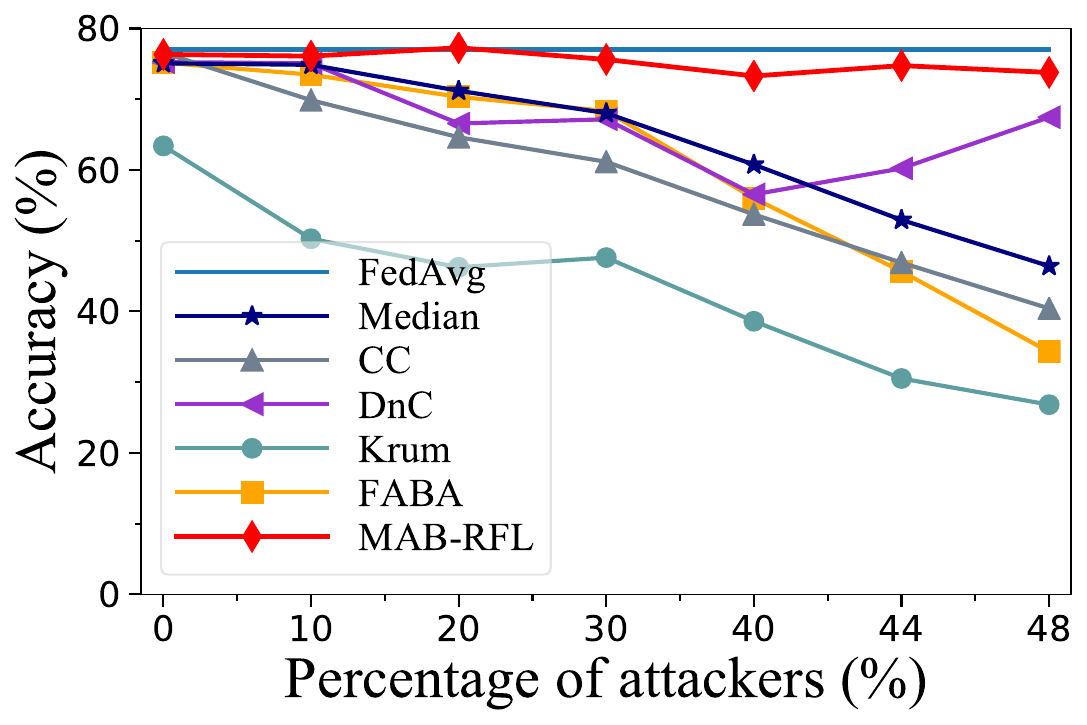}}
	\caption{Model accuracy under label flipping attack}
 \vspace{-5mm}
	\label{fig:lf}
\end{figure}

\begin{figure}[t]
	\centering
	\subfigure[MNIST]{
		\includegraphics[width=0.48\columnwidth]{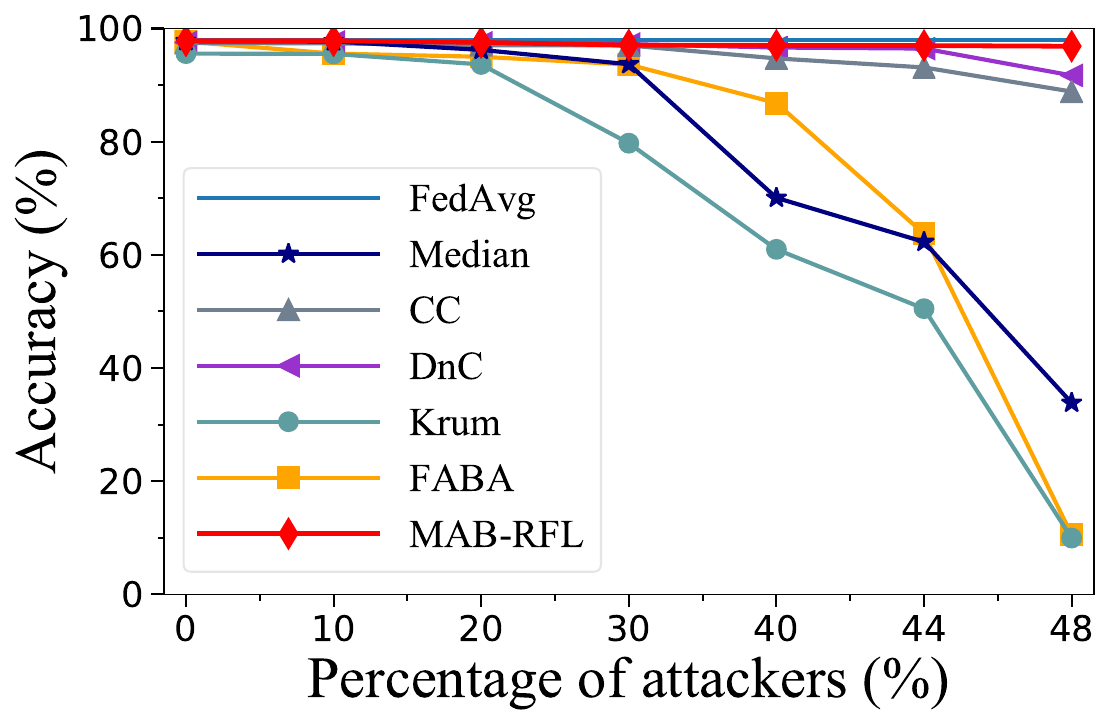}}
	\subfigure[CIFAR-10]{
		\includegraphics[width=0.48\columnwidth]{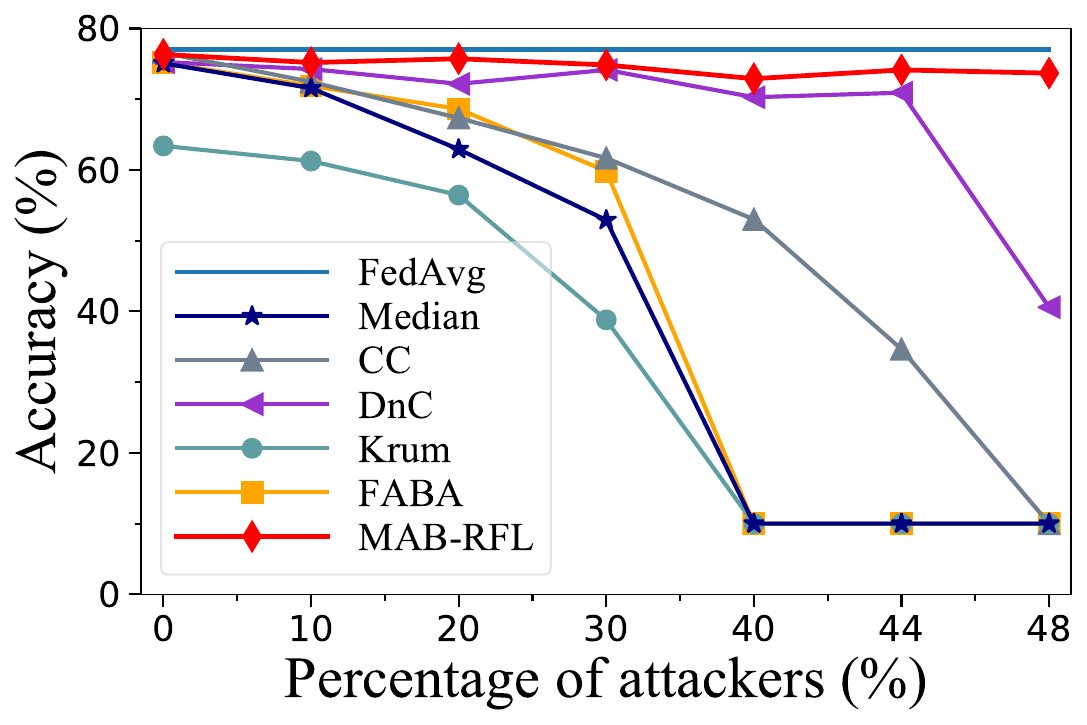}}
	\caption{Model accuracy under LIE attack}
  \vspace{-5mm}
	\label{fig:lie}
\end{figure}

\begin{figure}[t]
	\centering
	\subfigure[MNIST]{
		\includegraphics[width=0.48\columnwidth]{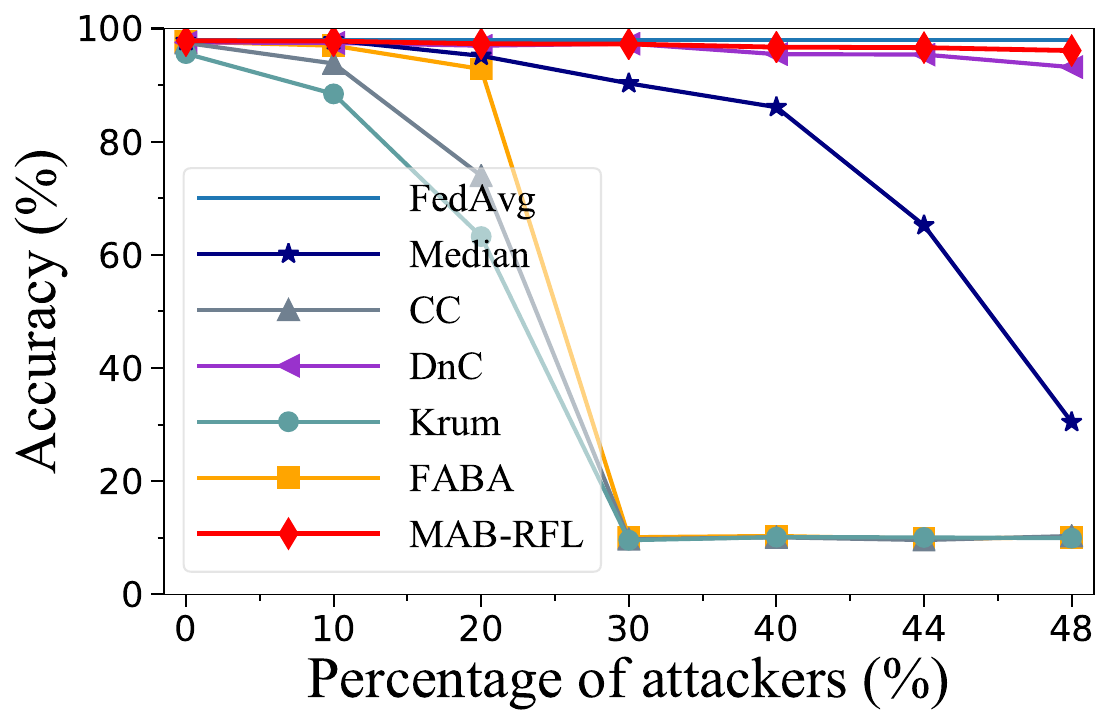}}
	\subfigure[CIFAR-10]{
		\includegraphics[width=0.48\columnwidth]{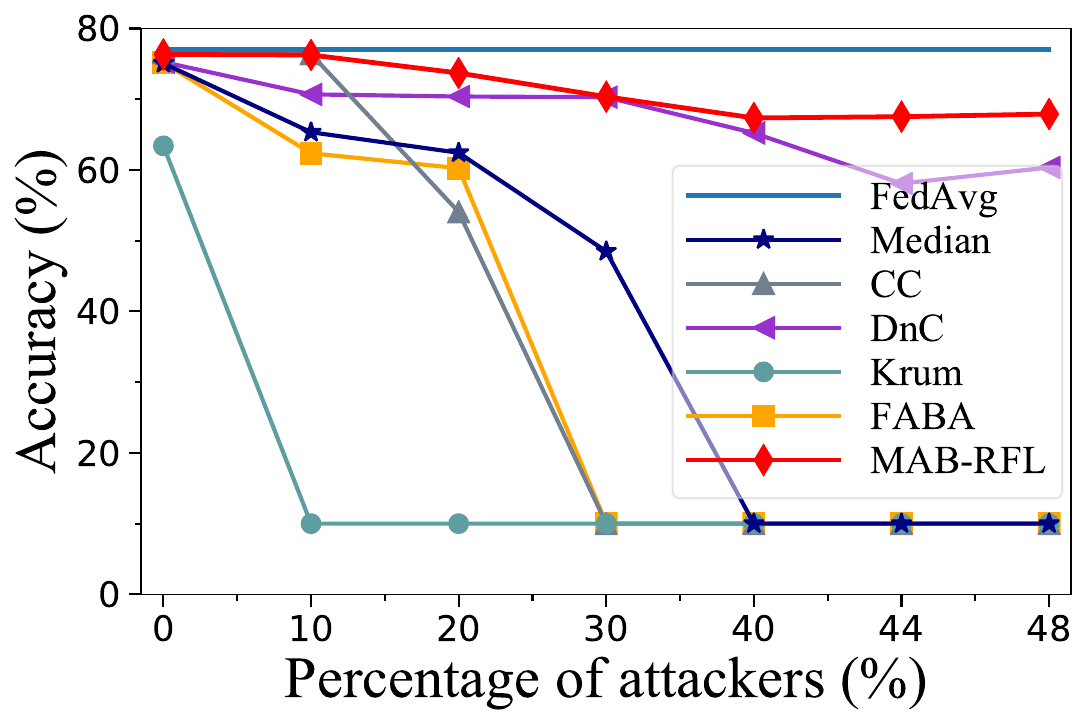}}
	\caption{Model accuracy under AGRT attack}
  \vspace{-5mm}
	\label{fig:agrt}
\end{figure}

\section{Experiments}
\subsection{Experimental Setup}
\paragraph{Datasets and Models.}
We evaluate MAB-RFL on MNIST and CIFAR-10. For MNIST, we use a \textit{convolutional neural network} (CNN) with 2 convolutional layers followed by 2 fully connected layers as the global model. For CIFAR-10, we employ a simplified version of VGG16 as the global model.

\paragraph{Data Distribution.}
Following previous works~\cite{FLTrust,AGRTailored}, we assign each client a random dominated label $l$. Then each training sample with label $l$ is distributed to the client with probability $q \in [0,1]$ and any other label with probability $\frac{1-q}{L-1}$, where $L$ is the total kinds of labels and $q$ is the non-IID degree. A larger $q$ indicates a higher degree of non-IID. Note that $q=\frac{1}{L}$ corresponds to the IID scenario. In our experiments, we set $q=0.5$ by default, which is the highest non-IID degree existing works considered. We assign different training set sizes for clients. Specifically, the training set size of each client is randomly chosen from $10$ to $500$ for MNIST and from $1200$ to $2000$  for CIFAR-10. 

{\color{black}\paragraph{Evaluated Attacks.}
We consider \textit{label flipping} (LF) attack~\cite{PCA}, \textit{little is enough} (LIE) attack~\cite{ALittleIsEnough}, and \textit{AGR-tailored} (AGRT) attack~\cite{AGRTailored}, which belong to agnostic attack, partial knowledge attack, and full knowledge attack, respectively.

\textbf{LF attack.} For each controlled client, the adversary flips the local sample labels  to generate misleading update. Specifically, the adversary flips the label $l$ of each training sample to $L-l-1$, where $L$ is the the number of classes in the classification problem and $l=0,1,\dots,L-1$.

\textbf{LIE attack.} The adversary first estimates coordinate-wise mean $\mu_{j}$ and deviation $\sigma_{j}$ through benign updates. Then the malicious update $\Tilde{g}$ can be crafted as follow:
\begin{equation}
	(\Tilde{g})_{j} = \mu_{j} - z^{max}\cdot\sigma_{j}, j\in[d],
\end{equation}
where $d$ is the total dimension of model update, and the positive attack factor $z^{max}$ determines the amount of noise added. A large $z^{max}$ makes the constructed update harmful to the global model but easy to be detected by the central server, while a small one is covert but less baleful. The original paper~\cite{ALittleIsEnough} advises that $z^{max}$ can be obtained by the Cumulative Standard Normal Function $\phi(z)$:
\begin{equation}
z^{max}=\max _{z}\left(\phi(z)<\frac{K-\left\lfloor\frac{K}{2}+1\right\rfloor}{m}\right).
\end{equation}

Note that all the controlled clients upload identical malicious update $\Tilde{g}$ to the central server, making the attack inconspicuous.

\textbf{AGRT attack.} AGRT attack aims to maximize the damage to the global model in each FL iteration in an optimized manner. Formally, the adversary performs the following optimization problem to construct malicious updates:
\begin{equation}
\begin{aligned}
\underset{\gamma}{\operatorname{argmax}} &\left\|\hat{g}-\mathscr{A}\left(g_{\{i \in[m]\}} \cup \widetilde{g}_{\{i \in[m+1,K]\}}\right)\right\| \\
& \widetilde{g}_{\{i \in[m+1,K]\}}=\hat{g}+\gamma \cdot \widetilde{\Delta} ; \hat{g}=\operatorname{FedAvg}\left(g_{\{i \in[m]\}}\right),
\end{aligned}
\end{equation}
where $\mathscr{A}$ is the known defense method (\ie, Krum) applied by the central server, $g_{\{i \in[m]\}}$ are the benign updates that the adversary knows (without loss of generality, we assume the first $m$ clients are benign and the rest are controlled by the adversary), $\hat{g}$ is a reference benign aggregation obtained by FedAvg~\cite{FedAvg} that averages all the benign updates that the adversary knows, $\widetilde{\Delta}$ is a perturbation vector, $\gamma$ is a scaling coefficient, $\widetilde{g}_{\{i \in[m+1,K]\}}$ are the final malicious updates.

It is worth noting that the above optimization incurs totally identical malicious updates for all controlled clients, thus MAB-RFL can easily discard all the malicious updates through Alg.~\ref{alg:sybil detection}. Nevertheless, considering that the adversary has the knowledge of the defense scheme in advance, we set the updates for half of the controlled clients to $\widetilde{g}$ and the other half to be the same as any benign update. This ensures that half of the malicious updates can circumvent the sybil detection and that one benign update is removed.

\paragraph{Evaluated Defenses.}
We compare MAB-RFL with three classical defenses, \ie, Krum~\cite{Krum}, FABA~\cite{FABA}, and Median~\cite{TrimmedMean}, as well as two newly proposed defenses, \ie, DnC~\cite{AGRTailored} and \textit{Centered Clip} (CC)~\cite{CC}.
 
\textbf{Krum.} Krum calculates the Euclidean distance between any two local gradients and selects the one that is closest to the $m-2$ neighboring local gradients.

\textbf{FABA.} FABA removes the local update that is farthest from the average of the local updates repeatedly until the number of eliminated updates reaches a predefined threshold.

\textbf{Median.} Median directly takes the coordinate-wise median value in each dimension of all local updates as the new global update.

\textbf{DnC.} DnC leverages \textit{singular value decomposition} (SVD) to extract the common features between benign and poisoned updates, and randomly samples a subset of parameters of each local update as its substitution, which will be projected along their top right singular eigenvector. Then an outlier score is obtained by computing the inner product of substitution and the projection, and the $K-m$ local updates with the highest scores will be removed.

\textbf{CC.} CC clips the local updates with large magnitudes, with the intuition that attackers may upload such updates to dominate the global model.
 
\paragraph{Evaluation Metrics and Parameter Setting.}
We use  \textit{accuracy}, which is the proportion of correctly predicted testing samples to total testing samples, to evaluate the final global model. We set the number of clients $K = 50$ for both datasets. To reduce the total communication rounds between clients and the server, we set the local epoch of each client to be 3. The total iteration $T = 100$. The importance of historical information $\lambda = 0.1$. For MNIST, we set the estimated maximum cosine similarity $c_{max} = 0.7$, minimum cosine similarity $c_{min} = 0.3$, and the acceptable difference between clusters $\alpha = -0.1$. For CIFAR-10, we set $c_{max} = 0.3$, $c_{min} = 0.1$, $\alpha = 0$. For AGRT attack, we choose $\widetilde{\Delta}$ as $-std\left(g_{\{i \in[m]\}}\right)$, \ie, the inverse standard deviation.
}

\paragraph{Comparison with the Existing Defenses.}
MAB-RFL achieves the three defense goals  as shown in Table~\ref{table:comparison}.

\textbf{Rationality: }MAB-RFL performs sampling with a more sophisticated method (\ie, rational sampling ) than the naive methods used in previous works, which improves the probability of selecting  benign clients. The toy example in section~\ref{Sec4.1: Motivation} well justifies this.

\textbf{Robustness: }MAB-RFL achieves the robustness objective, \ie, shielding FL from adversaries. As shown in Table~\ref{table:comparison}, MAB-RFL outperforms the existing defenses on three strong attacks over two non-IID dataset. For instance, in CIFAR-10, all the attacks pose no impact on MAB-RFL.

\textbf{Fidelity: }The accuracy of MAB-RFL under all attacks is always comparable to that of baseline (\ie, FedAvg) on two datasets, while the accuracy of Krum and Median are reducted even without attackers, which reflects that MAB-RFL is applicable for non-adversarial scenarios.

\paragraph{Impact of the Percentage of Attackers.}

Fig.~\ref{fig:lf}, Fig.~\ref{fig:lie}, and Fig.~\ref{fig:agrt} show that MAB-RFL always performs better or similar compared with state-of-the-art defenses in all settings. CC, FABA, and Median perform slightly worse than MAB-RFL when there are no more than $30\%$ attackers, however, their accuracy drops dramatically with $40\%$ attackers. DnC performs much better, but its accuracy drops obviously when there are nearly $50\%$ attackers. 


\begin{table}[!t]
\renewcommand\arraystretch{1}
\centering
\resizebox{\linewidth}{!}{
\begin{tabular}{|c|c|c||ccccccc|}
\hline
\multirow{2}{*}{\begin{tabular}[c]{@{}c@{}}Dataset\\ (Model)\end{tabular}} & \multirow{2}{*}{\begin{tabular}[c]{@{}c@{}}Percentage\\ of Attackers\end{tabular}} & \multirow{2}{*}{\begin{tabular}[c]{@{}c@{}}Attack \\ Type\end{tabular}} & \multicolumn{7}{c|}{Accuracy (\%)} \\ \cline{4-10} 
 &  &  & CC & DnC & FABA & Krum & Median & MAB-RFL & FedAvg \\ \hline\hline
\multirow{19}{*}{\begin{tabular}[c]{@{}c@{}}CIFAR-10\\ (VGG16)\end{tabular}} & 0\% & --- & \textbf{76.57} & 75.25 & 75.17 & 63.40 & 75.07 & 76.33 & 77.02 \\ \cline{2-10} 
 & \multirow{3}{*}{10\%} & AGRT & \textbf{76.39} & 70.68 & 62.36 & 10.00 & 65.33 & 76.25 & --- \\ \cline{3-10} 
 &  & LIE & 72.45 & 74.26 & 71.87 & 61.26 & 71.56 & \textbf{75.18} & --- \\ \cline{3-10} 
 &  & LF & 69.86 & 75.07 & 73.47 & 50.27 & 74.86 & \textbf{76.09} & --- \\ \cline{2-10} 
 & \multirow{3}{*}{20\%} & AGRT & 54.06 & 70.38 & 60.25 & 10.00 & 62.43 & \textbf{73.70} & --- \\ \cline{3-10} 
 &  & LIE & 67.34 & 72.17 & 68.62 & 56.45 & 62.93 & \textbf{75.72} & --- \\ \cline{3-10} 
 &  & LF & 64.62 & 66.57 & 70.33 & 46.25 & 71.19 & \textbf{77.27} & --- \\ \cline{2-10} 
 & \multirow{3}{*}{30\%} & AGRT & 10.00 & 70.28 & 10.00 & 10.00 & 48.49 & \textbf{70.35} & --- \\ \cline{3-10} 
 &  & LIE & 61.67 & 74.14 & 59.83 & 38.81 & 52.91 & \textbf{74.84} & --- \\ \cline{3-10} 
 &  & LF & 61.15 & 67.16 & 68.33 & 47.61 & 68.03 & \textbf{75.61} & --- \\ \cline{2-10} 
 & \multirow{3}{*}{40\%} & AGRT & 10.00 & 65.22 & 10.00 & 10.00 & 10.00 & \textbf{67.35} & --- \\ \cline{3-10} 
 &  & LIE & 53.01 & 70.28 & 10.00 & 10.00 & 10.00 & \textbf{72.90} & --- \\ \cline{3-10} 
 &  & LF & 53.71 & 56.54 & 56.00 & 38.61 & 60.75 & \textbf{73.28} & --- \\ \cline{2-10} 
 & \multirow{3}{*}{44\%} & AGRT & 10.00 & 58.08 & 10.00 & 10.00 & 10.00 & \textbf{67.53} & --- \\ \cline{3-10} 
 &  & LIE & 34.67 & 70.92 & 10.00 & 10.00 & 10.00 & \textbf{74.14} & --- \\ \cline{3-10} 
 &  & LF & 46.89 & 60.22 & 45.66 & 30.52 & 52.93 & \textbf{74.75} & --- \\ \cline{2-10} 
 & \multirow{3}{*}{48\%} & AGRT & 10.00 & 60.35 & 10.00 & 10.00 & 10.00 & \textbf{67.90} & --- \\ \cline{3-10} 
 &  & LIE & 10.00 & 40.59 & 10.00 & 10.00 & 10.00 & \textbf{73.68} & --- \\ \cline{3-10} 
 &  & LF & 40.42 & 67.45 & 34.29 & 26.82 & 46.41 & \textbf{73.78} & --- \\ \hline\hline
\multirow{19}{*}{\begin{tabular}[c]{@{}c@{}}MNIST\\ (CNN)\end{tabular}} & 0\% & --- & 97.45 & 97.53 & 97.68 & 95.56 & 97.76 & \textbf{97.82} & 98.01 \\ \cline{2-10} 
 & \multirow{3}{*}{10\%} & AGRT & 93.81 & 97.44 & 96.96 & 88.44 & \textbf{97.89} & 97.77 & --- \\ \cline{3-10} 
 &  & LIE & 97.36 & 97.58 & 95.56 & 95.50 & 97.70 & \textbf{97.83} & --- \\ \cline{3-10} 
 &  & LF & 95.51 & \textbf{97.65} & 95..61 & 67.41 & 96.03 & 97.11 & --- \\ \cline{2-10} 
 & \multirow{3}{*}{20\%} & AGRT & 73.97 & 96.96 & 92.83 & 63.23 & 95.16 & \textbf{97.29} & --- \\ \cline{3-10} 
 &  & LIE & 97.05 & 97.43 & 95.00 & 93.64 & 96.21 & \textbf{97.54} & --- \\ \cline{3-10} 
 &  & LF & 91.09 & \textbf{96.94} & 95.06 & 65.87 & 96.36 & 96.68 & --- \\ \cline{2-10} 
 & \multirow{3}{*}{30\%} & AGRT & 9.74 & \textbf{97.32} & 10.10 & 9.58 & 90.30 & 97.24 & --- \\ \cline{3-10} 
 &  & LIE & 97.00 & \textbf{97.23} & 93.63 & 79.73 & 93.68 & 97.05 & --- \\ \cline{3-10} 
 &  & LF & 83.77 & 96.64 & 94.00 & 59.45 & 95.09 & \textbf{97.10} & --- \\ \cline{2-10} 
 & \multirow{3}{*}{40\%} & AGRT & 10.09 & 95.46 & 10.32 & 10.12 & 86.09 & \textbf{96.70} & --- \\ \cline{3-10} 
 &  & LIE & 94.71 & 96.63 & 86.82 & 60.97 & 70.07 & \textbf{97.01} & --- \\ \cline{3-10} 
 &  & LF & 69.57 & 92.83 & 92.27 & 45.49 & 92.05 & \textbf{97.03} & --- \\ \cline{2-10} 
 & \multirow{3}{*}{44\%} & AGRT & 9.61 & 95.39 & 9.82 & 10.07 & 65.21 & \textbf{96.61} & --- \\ \cline{3-10} 
 &  & LIE & 93.11 & 96.45 & 63.73 & 50.49 & 62.28 & \textbf{96.97} & --- \\ \cline{3-10} 
 &  & LF & 58.12 & 89.39 & 70.37 & 50.37 & 80.32 & \textbf{96.48} & --- \\ \cline{2-10} 
 & \multirow{3}{*}{48\%} & AGRT & 10.32 & 93.16 & 10.09 & 9.93 & 30.44 & \textbf{96.09} & --- \\ \cline{3-10} 
 &  & LIE & 88.85 & 91.70 & 10.51 & 9.96 & 33.85 & \textbf{96.86} & --- \\ \cline{3-10} 
 &  & LF & 48.10 & 77.91 & 63.94 & 3.68 & 64.47 & \textbf{95.26} & --- \\ \hline
\end{tabular}
}
\caption{A comparison with existing defense methods on three attacks over two non-IID datasets}
\label{table:comparison}
\end{table}

\section{Conclusion}
We propose a new robust federated learning scheme to make the server perform rational client selection decisions. Two approaches are designed to identify malicious updates from sybil and non-sybil attacks, based on which rewards for each client selection decision can be accurately evaluated.   Extensive experimental results show that MAB-RFL outperforms existing defenses.

\section*{Acknowledgments}
Shengshan’s work is supported in part by the National Natural Science Foundation of China (Grant No. U20A20177). Leo’s work is supported in part by the National Natural Science Foundation of China (Grant No. 61702221). Yuanyuan’s work is supported in part by the National Natural Science Foundation of China (Grant No. 62002127).

\bibliographystyle{named}
\bibliography{ijcai22}

\clearpage
\appendix 


\end{document}